\documentstyle[12pt]{article}
\def\abstracts#1#2#3{{
        \centering{\begin{minipage}{4.62in}\baselineskip=13pt
        \small
        \centerline{\bf Abstract}
        \vspace*{0.2cm}                % W. Janke (July 1, 1992)
        \parindent=0pt #1\par
        \parindent=18pt #2\par
        \parindent=15pt #3
        \end{minipage} }\par}}
\renewcommand{\thefootnote}{\fnsymbol{footnote}}
\begin{document}
\vspace*{-3cm}
\hfill \parbox{4.2cm}{ preprint UAB-FT-318\\
                       hep-lat/9310025\\
                                         }\\[1.5cm]
%\vspace*{1.5cm}
\centerline{\LARGE \bf Ising Model Universality for}\\[0.2cm]
\centerline{\LARGE \bf Two-Dimensional
Lattices\footnotemark}\\[0.6cm]
\footnotetext{\noindent Work supported in part by The Florida High
Technology and Industry Council under\\ \mbox{~~~~~~}Contract FHTIC-15423ERAU.}
\addtocounter{footnote}{-1}
\renewcommand{\thefootnote}{\arabic{footnote}}
%\vspace*{0.4cm}
\centerline{\large {\em Wolfhard Janke\/}$^{1}$,
                   {\em Mohammad Katoot\/}$^{2}$
             and   {\em Ramon Villanova\/}$^{2,3}$}\\[0.4cm]
\centerline{\large    $^1$ {\small Institut f\"ur Physik,
                      Johannes Gutenberg-Universit\"at Mainz}}
\centerline{    {\small 55099 Mainz, Germany }}\\[0.15cm]
\centerline{\large    $^2$ {\small Department of Physical
                      Sciences,
                      Embry-Riddle Aeronautical University}}
\centerline{    {\small Daytona Beach, Florida 32114, USA}}\\[0.15cm]
\centerline{\large    $^3$ {\small Grup de F\'{\i}sica Te\`{o}rica and IFAE,
            Facultat de Ci\`{e}ncies, Universitat Aut\`{o}noma de Barcelona }}
\centerline{    {\small 08193 Bellaterra, Spain}}\\[2.50cm]
%supported by a fellowship
%  from the "Centre de Supercomputaci\'{o} de Catalunya".

%\vspace*{2.3cm}
\abstracts{}{
We use the single-cluster Monte Carlo update algorithm to simulate the
Ising model on two-dimensional Poissonian random lattices of Delaunay type
with up to 80\,000 sites. By applying reweighting techniques and finite-size
scaling analyses to time-series data near criticality, we obtain unambiguous
support that the critical exponents for
the random lattice agree with the exactly known exponents for regular lattices,
i.e., that (lattice) universality holds for the two-dimensional Ising model.
}{}
\thispagestyle{empty}
\newpage
\pagenumbering{arabic}
%
%-------------------------------------------------------------------
     \section{Introduction}
%-------------------------------------------------------------------
%
In numerical simulations of many physical systems
random lattices \cite{mei,cfl} are a useful tool to discretize space
without introducing any kind of anisotropy. Recent applications of various
types
of random lattices can be found in a great variety of fields, such as
quantum field theory or quantum gravity \cite{cfl,qft,ren},
the statistical mechanics of membranes \cite{membranes},
diffusion limited aggregation \cite{dla}, or
growth models of sandpiles \cite{puhl},
to mention a few. As a consequence of
the preserved rotational (or more generally, Poincar\'e) invariance,
spin systems or field theories defined on random lattices are expected to
reach the infinite volume or continuum limit faster than on regular lattices.
Implicit in this approach is the assumption of (lattice) universality which
states that systems defined on lattices of different type should exhibit the
same qualitative behaviour once the physical length scale is much larger
than the average lattice spacing. While this assumption is known to be
true for spin systems on different {\em regular} lattices, previous numerical
work \cite{all,espriu} on random lattices could only give weak evidence that
universality holds in this case as well.

In this note we reconsider the Ising model defined on two-dimensional
Poissonian\footnote{For alternative site distributions see, e.g.,
Refs.\cite{dla,caer}.} random
lattices constructed according to the Voronoi/Delaunay prescription
\cite{cfl,ren}.
In previous work on this model, Espriu {\em et al.} \cite{espriu} have used
standard Metropolis Monte Carlo (MC) simulations on lattices with $N=10\,000$
sites to study the approach of criticality in the low- and high-temperature
phase. Here we report high-statistics simulations in the very vicinity of the
phase transition, using considerably larger lattices of size up to $N=80\,000$.
To achieve the desired accuracy of the
data we made extensively use of recently developed greatly refined MC
simulation techniques, such as cluster update algorithms \cite{clu1,clu2}
and reweighting methods \cite{reweight}. As a result of finite-size
scaling (FSS) analyses of our data
we obtain very strong support for (lattice) universality in this model.
\section{Model}
As partition function we take
\begin{equation}
Z = \sum_{\{s_i\}} e^{-K E};\mbox{~~} E = -\sum_{\langle ij \rangle} s_i s_j;
\mbox{~~} s_i=\pm 1,
\label{eq:0}
\end{equation}
where $K= J/k_BT > 0$ is the inverse temperature in natural units and
$\langle ij \rangle$ denote nearest-neighbour links of the Delaunay random
lattices, computed according to the (dual) Voronoi cell construction as
described, e.g., in Ref.\cite{ren}.
Following Ref.\cite{espriu} we thus take the relative weights of the links
to be constant.
The lattice sizes studied are $N=5\,000$, $10\,000$,
$20\,000$, $40\,000$, and $80\,000$, with three replicas for each of the two
smallest
lattices, and two replicas for $N=20\,000$. We always employed periodic
boundary
conditions, i.e., the topology of a torus. In this case Euler's theorem
implies $\overline{q}=6$, where $\overline{q}$ is the lattice average of
the local coordination numbers $q$ that vary for Poissonian random lattices
between $3$ and $\infty$.
All our lattices satisfy this rigorous constraint, and also the distributions
$P(q)$ agree well with numerical evaluations of exact integral expressions
\cite{drouffe}. The highest coordination number we actually observed in our
simulations was $q=13$ in the $N=80\,000$ lattice.

To update the spins $s_i$ we employed the single-cluster update
algorithm \cite{clu2} which is straightforward to adopt to random lattices.
 From comparative studies \cite{compare} on regular lattices the single-cluster
update is
expected to be more efficient than the multiple cluster variant \cite{clu1}.
All runs
were performed at $K=0.263$, the estimate of the critical coupling $K_c$ as
quoted by Espriu {\em et al.} \cite{espriu}. After discarding from
$50\,000$ to
$150\,000$ clusters to reach equilibrium from an initially completely
disordered state, we generated a further
$4 \times 10^6$ clusters and recorded every $10th$ cluster measurements of
the energy per spin, $e=E/N$, and the magnetization per spin, $m=\sum_i s_i/N$
in a time-series file.
 From analyses of the autocorrelation functions of $e$ and
$m^2$ we obtained at the scale of our measurements integrated
autocorrelation times of
$\hat{\tau}_e \approx 0.8 - 1.3$ and $\hat{\tau}_{m^2} \approx 0.7 - 0.9$,
respectively. Our samples thus consist effectively of about $200\,000$
statistically independent data. The statistical errors are estimated by
deviding the time series into 20 blocks, which are jack-knived to avoid
bias problems in reweighted data.
\section{Results}
To determine the transition point $K_c$ and the correlation length exponent
$\nu$ we first concentrated on the magnetic Binder parameter \cite{binder},
\begin{equation}
    U_L(K) = 1 - \frac{\langle m^4 \rangle}{3 \langle m^2 \rangle^2},
\label{eq:1}
\end{equation}
where $L \equiv \sqrt{N}$ is defined as the linear length of the lattice
in natural units. It is well known \cite{binder} that
the curves $U_L(K)$ for lattices of size $L$ and $L'$ should intersect in
points $(K^{\times}(L,L'), U^{\times}(L,L'))$ which approach $(K_c,U^*)$ for
large $L$,$L'$, and the derivatives $U'_L \equiv dU_L/dK$ at these points
should scale
asymptotically with $L^{1/\nu}$. Our results for $U_L(K)$
obtained from reweighting the time-series data at $K=0.263$ are plotted in
Fig.~1. For the small lattices the curves are an average over the different
replicas \cite{tobe}.

Taking as estimate for $K_c$ the average of the $K^{\times}(L,L')$ for the
three largest lattices, we obtain
\begin{equation}
K_c = 0.2630 \pm 0.0002,
\label{eq:2}
\end{equation}
where the (rough) error estimate reflects also the fluctuations between
different replicas. The value (\ref{eq:2}) is in very good agreement with
high-temperature series expansion analyses
($K_c \approx 0.26303$) \cite{espriu} and
MC simulations in the disordered phase ($K_c = 0.2631(3)$) \cite{espriu}.

At the critical coupling (\ref{eq:2}), $U_L(K)$ varies only little and an
average over all lattice sizes gives
\begin{equation}
U^* = 0.6123 \pm 0.0025.
\label{eq:3}
\end{equation}
At $K=K_c-0.0002$ and $K=K_c+0.0002$ we obtain
$U^* = 0.6054(25)$ and $U^* = 0.6183(28)$, respectively. The value (\ref{eq:3})
is in very good agreement with MC estimates for the regular simple square
({\em sq\/}) lattice which are $U^* = 0.615(10)$ \cite{landau} and
$U^* = 0.611(1)$ \cite{heermann}. This agreement may be taken as a first
indication of lattice universality.

To get an estimate for the exponent $\nu$ we have computed the effective
exponents
\begin{equation}
   \nu_{\rm eff} = \frac{ \ln(L'/L) }
              { \ln \left( U'_{L'}(K^{\times}) / U'_L(K^{\times}) \right) }
\label{eq:4}
\end{equation}
for all possible combinations of $L$ and $L'$. Since we do not observe any
definite trend of $\nu_{\rm eff}$ as a function of $L$ and $L'$, we quote as
our final result for $\nu$ the average over all combinations,
\begin{equation}
\nu = 1.008 \pm 0.022,
%\mbox{~~~~~ (effective $\nu$'s)},
\label{eq:5}
\end{equation}
where the error estimate is the standard deviation of the $\nu_{\rm eff}$.
If we consider only the crossing points with the $N=80\,000$ curve, the
estimate for $\nu$ even sharpens to $\nu = 1.0043 \pm 0.0036$. We can thus
conclude that our estimate of the exponent $\nu$ for the random lattice is
fully consistent with the exact regular lattice value of $\nu=1$.

The ratio of exponents $\gamma/\nu$ follows from the scaling of the maxima,
$\chi'_{\rm max}(L) \propto L^{\gamma/\nu}$, of the (finite lattice)
susceptibility
\begin{equation}
    \chi'(K) = K \, N ( \langle m^2 \rangle - \langle |m| \rangle^2 ).
\label{eq:6}
\end{equation}
The curves of $\chi'(K)$ obtained by reweighting the primary data at $K=0.263$
are shown in Fig.~2. It is then
straightforward to determine the maxima $\chi'_{\rm max}$ for each
lattice size $L$, and a straight line fit
through all data points in a log-log plot of $\chi'_{\rm max}$ vs $L$ gives
\begin{equation}
\gamma/\nu = 1.7503 \pm 0.0059,
\label{eq:7}
\end{equation}
with a goodness-of-fit parameter \cite{numlib} of $Q=0.035$. This is again
in perfect
agreement with the exact value for regular lattices, $\gamma/\nu=1.75$. We
can thus conclude that universality also holds as far as the exponent ratio
$\gamma/\nu$ is concerned.

The locations of the susceptibility maxima, $K^{\chi'}_{\rm max}$, should
scale for large $L$ according to $K^{\chi'}_{\rm max} = K_c + aL^{-1/\nu}$,
where $a$ is a
non-universal constant. Assuming $\nu=1$ and performing a linear fit through
the
$K^{\chi'}_{\rm max}$ of the three largest lattices we obtain
$K_c=0.262947(77)$
with $Q=0.24$, in good agreement with our earlier estimate from the
intersection
points of the parameter $U_L$.

Having estimated $\nu$ and $\gamma$, all other exponents can in principle
be calculated by scaling or hyperscaling relations, e.g.,
$2\beta/\nu = d - \gamma/\nu$, where $d$ is the dimension.
To get an independent estimate for the exponent ratio  $\beta/\nu$ we
have considered the FSS behaviour of the magnetization $\langle |m| \rangle$ at
its point of inflection, which is given by
$\langle |m| \rangle|_{\rm inf}(L) \propto L^{-\beta/\nu}$. From a linear fit
through all data points we obtain
\begin{equation}
\beta/\nu = 0.1208 \pm 0.0092,
\end{equation}
with $Q=0.10$. Also this result is perfectly compatible with the exact
value for regular lattices, $\beta/\nu = 0.125$, thus supporting the
hyperscaling hypothesis for random lattices as well.

Furthermore, from the asymptotic scaling of the points of inflection,
$K^{\langle |m| \rangle}_{\rm inf} = K_c + a'L^{-1/\nu}$, we can get
another estimate for the critical coupling. Assuming again $\nu=1$, a fit
through the points of the three largest lattices yields $K_c = 0.26304(14)$
with $Q=0.60$, thus confirming our previous estimates.

Let us finally consider the specific heat,
\begin{equation}
C = K^2 N ( \langle e^2 \rangle - \langle e \rangle^2),
\label{eq:8}
\end{equation}
and the associated critical exponent $\alpha$. Here hyperscaling predicts
$\alpha = 2 - d\nu$. Since we already know that $\nu \approx 1$
we thus expect $\alpha \approx 0$ for two-dimensional random lattices. The
corresponding FSS prediction for the maxima of $C$ is then
\begin{equation}
C_{\rm max}(L) =  B_0 + B_1 \ln L,
\label{eq:9}
\end{equation}
with non-universal constants $B_0$ and $B_1$.
The semi-log plot in Fig.~3 clearly demonstrates that our data is
consistent with this prediction. A linear fit through all data points gives
$B_0 = 0.346(52)$ and $B_1 = 0.391(12)$ with $Q=0.84$. On the other hand, we
cannot claim unambiguous support for logarithmic scaling. In fact, we can even
fit the data with a pure power-law Ansatz, $C_{\rm max} \propto
L^{\alpha/\nu}$,
yielding $\alpha/\nu = 0.1824(53)$ with a similar goodness-of-fit
parameter, $Q=0.93$, as for the logarithmic fit. We also tried a non-linear
three-parameter fit to the more reasonable Ansatz
$C_{\rm max} = b_0 + b_1 L^{\alpha/\nu}$. Even though the exponent ratio
$\alpha/\nu = 0.17(16)$ then comes out consistent with zero, the errors on all
three parameters are much too large to draw a firm conclusion from such a fit.
By means of exact results for the {\em sq\/} lattice \cite{ferdinand},
we have checked \cite{tobe} that for the regular lattice the specific heat
behaves very similar. In both cases one would need much
larger lattice sizes to discriminate between logarithmic and power-law scaling.

As before the peak locations $K^C_{\rm max}$ should scale like
$K^C_{\rm max} = K_c + a'' L^{-1/\nu}$. Assuming again $\nu=1$, we obtain from
a fit to the data for the three largest lattices $K_c = 0.26295(33)$ with
$Q=0.95$, in agreement with the previous estimates.
\section{Conclusion}
In summary, we have performed a fairly detailed finite-size scaling study of
the Ising model on two-dimensional Poissonian random lattices of the Delaunay
type. Our estimate for the critical coupling derived from the intersection
points of the Binder parameter is $K_c=0.2630(2)$, the inflection points of
the magnetization yield asymptotically $K_c=0.26304(14)$, and from the peak
locations of the suceptibility and specific heat we extrapolate
$K_c=0.262947(76)$ and $K_c=0.26295(33)$, respectively. These values are in
good
agreement with previous simulations in the disordered phase and with analyses
of high-temperature series expansions by Espriu {et al.\/} \cite{espriu}.

As usual the specific-heat maxima are difficult to analyze, since the
asymptotic finite-size scaling behaviour sets in only for extremely large
lattice sizes. Our data is consistent with a logarithmic scaling, i.e., with a
critical exponent $\alpha=0$, but not yet sufficient to exclude a power-law
scaling with $\alpha \ne 0$ on a statistically firm basis. Precisely the same
situation is encountered, however, for the (exactly known) specific heat
of the regular {\em sq\/} lattice. We take this observation as further support
that also for this quantity there is no violation of universality.

Our results for the critical exponents $\nu$, $\gamma$ and $\beta$ are much
easier to interpret. They clearly
indicate that these exponents have the same values as for regular
lattices, i.e., here we obtain strong support for lattice universality in the
two-dimensional Ising model.

As a future project it would be interesting to repeat this study for
{\em dynamical\/}
random lattices that satisfy the Voronoi/Delaunay construction at all times
\cite{dynvorlat}. The important question would be whether the critical
behaviour
is still governed by the critical exponents of the {\em static\/} random (or
regular) lattice considered here, or by the critical exponents predicted by
matrix model theory \cite{matrix}. For standard dynamically triangulated
lattices, which do {\em not\/} satisfy
the Voronoi/Delaunay construction, strong numerical evidence for the second
alternative was reported recently in Ref.\cite{ben-av}.
\section*{Acknowledgements}
R.V. is supported by a fellowship
from the ``Centre de Supercomputaci\'{o} de Catalunya'', and W.J.
thanks the Deutsche Forschungsgemeinschaft for a Heisenberg fellowship.
Some of the simulations were performed on the SCRI cluster of fast
RISC workstations.
\clearpage
\newpage
%
%-----------------------------------------------------------------------
     
\newpage
%
%----------------------------------------------------------------
     {\Large\bf Figure Headings}
%----------------------------------------------------------------
%
  \vspace{1in}
  \begin{description}
    \item[\tt\bf Fig. 1:]
The parameter $U_L(K)$ vs the inverse temperature $K$ for random lattices
of size $L=\sqrt{N}$ with $N=5\,000, 10\,000, 20\,000, 40\,000$ and
$80\,000$. The curves are obtained by reweighting the time-series data
at $K=0.263$ ($\approx K_c$).
    \item[\tt\bf Fig. 2:]
The (finite lattice) susceptibility $\chi'(K)$ for the same random lattices
as in Fig.~1. The curves are obtained by reweighting the time-series data at
$K=0.263$ ($\approx K_c$).
   \item[\tt\bf Fig. 3:]
Finite-size scaling plot of the specific-heat maxima $C_{\rm max}$ vs
$\ln L$, where $L=\sqrt{N}$. The
solid straight line shows the least-squares fit
$C_{\rm max} = B_0 + B_1 \ln L$, with $B_0 = 0.346(52)$ and $B_1 = 0.391(12)$.
 \end{description}
\end{document}